\documentstyle[bezier,12pt]{article}
\makeatletter
\newdimen\normalarrayskip              
\newdimen\minarrayskip                 
\normalarrayskip\baselineskip
\minarrayskip\jot
\newif\ifold             \oldfalse
\newif\ifdisplayarray    \displayarraytrue
\newif\ifbigarray        \bigarraytrue
\def\arraymode{\ifold\relax\else\ifdisplayarray\displaystyle\else\relax\fi\fi}
\def\eqnumphantom{\phantom{(\theequation)}}     
\def\@arrayskip{\ifold\baselineskip\z@\lineskip\z@\else\ifbigarray
     \baselineskip\normalarrayskip\lineskip\minarrayskip
     \else
     \baselineskip\z@\lineskip\z@\fi\fi}
\def\@arrayclassz{\ifcase \@lastchclass \@acolampacol \or
\@ampacol \or \or \or \@addamp \or
   \@acolampacol \or \@firstampfalse \@acol \fi
\edef\@preamble{\@preamble
  \ifcase \@chnum
     \hfil$\relax\arraymode\@sharp$\hfil
     \or $\relax\arraymode\@sharp$\hfil
     \or \hfil$\relax\arraymode\@sharp$\fi}}
\def\@array[#1]#2{\setbox\@arstrutbox=\hbox{\vrule
     height\arraystretch \ht\strutbox
     depth\arraystretch \dp\strutbox
     width\z@}\@mkpream{#2}\edef\@preamble{\halign \noexpand\@halignto
\bgroup \tabskip\z@ \@arstrut \@preamble \tabskip\z@ \cr}%
\let\@startpbox\@@startpbox \let\@endpbox\@@endpbox
  \if #1t\vtop \else \if#1b\vbox \else \vcenter \fi\fi
  \bgroup \let\par\relax
  \let\@sharp##\let\protect\relax
  \@arrayskip\@preamble}
\def\eqnarray{\stepcounter{equation}%
              \let\@currentlabel=\theequation
              \global\@eqnswtrue
              \global\@eqcnt\z@
              \tabskip\@centering
              \let\\=\@eqncr
              $$%
 \halign to \displaywidth\bgroup
    \eqnumphantom\@eqnsel\hskip\@centering
    $\displaystyle \tabskip\z@ {##}$%
    &\global\@eqcnt\@ne \hskip 2\arraycolsep
         \hfil$\arraymode{##}$\hfil
    &\global\@eqcnt\tw@ \hskip 2\arraycolsep
         $\displaystyle\tabskip\z@{##}$\hfil
         \tabskip\@centering
    &{##}\tabskip\z@\cr}
\newenvironment{marray}{\begin{equation}\begin{array}}%
{\end{array}\end{equation}}
\newenvironment{carray}{\begin{equation}\begin{array}{rcl}}%
{\end{array}\end{equation}}
\def\be{\@ifnextchar[{\def\ee{\end{equation}}\begin{equation}\l@b}%
{\def\ee{$$}$$}}
\def\l@b[#1]{\label{#1}}
\def\ba{\@ifnextchar[{\def\ee{\end{carray}}\begin{carray}\l@b}%
{\def\ee{\end{array}$$}$$\begin{array}{rcl}}}
\def\barray#1{\@ifnextchar[{\def\ee{\end{marray}}\begin{marray}{#1}\l@b}%
{\def\ee{\end{array}$$}$$\begin{array}{#1}}}
\def\herring{\@ifnextchar[{\@herring}{\@herring[\vcenter]}}
\def\@herring[#1]#2{\begingroup
\def\*{\\ \>}
\topsep0pt
\partopsep0pt
\def\tabbing{\lineskip\jot \lineskiplimit\jot
     \let\>\@rtab\let\<\@ltab\let\=\@settab
     \let\+\@tabplus\let\-\@tabminus\let\`\@tabrj\let\'\@tablab
     \let\\=\@tabcr
     \global\@hightab\@firsttab
     \global\@nxttabmar\@firsttab
     \dimen\@firsttab\@totalleftmargin
     \global\@tabpush0 \global\@rjfieldfalse
     \trivlist \item[]\if@minipage\else\vskip\parskip\fi
     \setbox\@tabfbox\hbox{\rlap{\indent\hskip\@totalleftmargin
       \the\everypar}}\def\@itemfudge{\box\@tabfbox}\@startline\ignorespaces}
\def\@startfield{\global\setbox\@curfield\hbox
                    \bgroup$\displaystyle}%
\def\@stopfield{$\egroup}%
#1{\begin{tabbing}#2\end{tabbing}}\endgroup}
\def\eq#1{(\ref{#1})}
\def\theequation{\thesection.\arabic{equation}}
\@addtoreset{equation}{section}%
\def\@cite#1#2{\hbox{ [#1\if@tempswa ,#2\fi]}}
\def\@citex[#1]#2{\if@filesw\immediate\write\@auxout{\string\citation{#2}}\fi
  \def\@citea{}\@cite{\@for\@citeb:=#2\do
    {\@citea\def\@citea{,\penalty\@m}\@ifundefined  
       {b@\@citeb}{{\bf ?}\@warning
       {Citation `\@citeb' on page \thepage \space undefined}}%
\hbox{\csname b@\@citeb\endcsname}}}{#1}}
\def\@sect#1#2#3#4#5#6[#7]#8{\ifnum #2>\c@secnumdepth
     \def\@svsec{}\else
     \refstepcounter{#1}\edef\@svsec{\csname the#1\endcsname.%
     \hskip 0.8em }\fi
     \@tempskipa #5\relax
      \ifdim \@tempskipa>\z@
        \begingroup #6\relax
          \@hangfrom{\hskip #3\relax\@svsec}{\interlinepenalty \@M #8\par}%
        \endgroup
       \csname #1mark\endcsname{#7}\addcontentsline
         {toc}{#1}{\ifnum #2>\c@secnumdepth \else
                      \protect\numberline{\csname the#1\endcsname}\fi
                    #7}\else
        \def\@svsechd{#6\hskip #3\@svsec #8\csname #1mark\endcsname
                      {#7}\addcontentsline
                           {toc}{#1}{\ifnum #2>\c@secnumdepth \else
                             \protect\numberline{\csname the#1\endcsname}\fi
                       #7}}\fi
     \@xsect{#5}}

\makeatother
\begin{document}
\def\e{\mbox{e}}
\def\noi{\noindent}
\def\sgn{{\rm sgn}}
\def\gsim{\;
\raise0.3ex\hbox{$>$\kern-0.75em\raise-1.1ex\hbox{$\sim$}}\; }
\def\lsim{\;
\raise0.3ex\hbox{$<$\kern-0.75em\raise-1.1ex\hbox{$\sim$}}\; }
\def\MeV{\rm MeV}
\def\eV{\rm eV}
\def\VEV#1{\left\langle #1\right\rangle}
\newcommand{\beqa}{\begin{eqnarray}}
\newcommand{\eeqa}{\end{eqnarray}}
\topskip -12mm

{\title{\vskip-3truecm{\hfill {{\small HU-TFT-95-28\\
        }}\vskip 1truecm}
{\bf Neutrino conversions in hot plasma}}


\author{
{\sc Kari Enqvist$^{1}$, Jukka Maalampi$^2$} \\
{\small\sl Department of  Physics,
 University of Helsinki, Finland}\\
and\\
{\sc V.B. Semikoz$^{3,4}$ }\\
{\small\sl  Department de Fisica Teorica,
 University of Valencia, Spain}
 \\
}
\maketitle}
\vspace{1cm}
\begin{abstract}
\noindent We discuss the excitation of sterile neutrinos in the
early universe using general quantum kinetic equations, which also
incorporate a possible primordial magnetic field $B$.
We find a new contribution
to the excitation propability, which has its origin in the shrinkage of
the spin vector. In the absence of $B$ nucleosynthesis implies the
constraint
$\mid \Delta m^2\mid\sin^22\theta_0\lsim 1.6\times 10^{-6}~{\rm eV}^2$
which is more restrictive than previous estimates. We also present
examples of possible stringent limits for $B\ne 0$.
\end{abstract}
\vfill
\footnoterule

\baselineskip=-1.5mm
{\small\noindent
$^1$enqvist@phcu.helsinki.fi;~ $^2$maalampi@phcu.helsinki.fi;~
$^3$semikoz@evalvx.ific.uv.es~; $^4$On a leave of absence from the
Institute for  Terrestrial Magnetism, Ionosphere and Radio Wave
Propagation, Russian Academy of Sciences, IZMIRAN,Troitsk,  Russia.}

\thispagestyle{empty}
\newpage
\setcounter{page}{1}
\baselineskip=8mm
\section{Introduction.} The recent observation of the cosmic
microwave background temperature  anisotropy on large scales by
the COBE satellite has hinted towards the existence  of a hot
component in the dark matter (HDM) of the universe
\cite{Smoot}. A light neutrino is an obvious candidate for hot
dark matter. As is well known, there are a number of unrelated astrophysical
and cosmological constraints on light neutrinos, which have to do with solar
neutrinos and the deficit of muon neutrinos in the
atmospheric neutrino fluxes\cite{solar}. Reconciling all
these issues in the context of a
three generation mixing model only is quite difficult, and as was
first pointed out in\cite{ektatmos}, if neutrinos are to solve
the dark matter, solar and atmospheric problems simultaneously,
one actually has to introduce a fourth neutrino, which could well
be a sterile neutrino. Moreover, it has been suggested\cite{subir}
that the recently
observed signals in the KARMEN detector could be explained by the
the decay pattern $\pi^+\to\mu^+\nu_s$, where $\nu_s$ is light sterile
neutrino.
Simple extensions of the standard
electroweak model which can accommodate all known hints for
neutrino masses, including solar and atmospheric neutrino
observations also postulate the existence of a light neutrino
$\nu_s$
\cite{Valle1,Valle2}. In some of these models such sterile
neutrino is the   HDM candidate\cite{Valle2}. (Sterile neutrino
as warm dark matter has recently been discussed in
\cite{dodelson}.)

Very tight constraints on the neutrino mass matrix that includes  a
singlet
$\nu_s$ can be deduced from the primordial nucleosynthesis bound
on the excess relic energy density at the proton-neutron
freeze-out, which takes place at $T\approx 0.7$ MeV. The bound
derives from the evaluations of the abundances of primordial
$^4$He and D and is usually quantified in units of relativistic
neutrino species. For a recent discussion on the neutrino bound, see\cite{OS}
(see also\cite{Schramm}).

Because of the mixing, sterile neutrinos can be produced in weak
collisions and thus be brought into equilibrium, whence they
would count as an extra relativistic neutrino degree of freedom,
in clear contradiction with the observations \cite{early}.
Equilibration can be  avoided for a certain range of the mixing
parameters, though. The region of the active-sterile
neutrino oscillation parameters $\Delta m^2= m^2_2 - m^2_1$ and
$\sin^2 2\theta_0$ excluded by the Big Bang Nucleosynthesis (BBN)
has been estimated to be\cite{EKT}
\be[isotropic]
\begin{array}{cc}
\mid \Delta m^2\mid \sin^42\theta_0 \gsim 5\times 10^{-6}~{\rm
eV}^2, ~~~\nu_a=\nu_e,\\

\mid \Delta m^2\mid \sin^42\theta_0 \gsim 3\times 10^{-6}~{\rm
eV}^2, ~~~\nu_a=
\nu_{\mu, \tau}.\\
\end{array}
\ee  Folding in an actual nucleosynthesis code  into neutrino
evolution equations has verified this result\cite{DS}.

Strong random magnetic fields  change neutrino oscillations
drastically. This is a pertinent issue because of the possibility
that the observed galactic magnetic fields, which are of the
order of microgauss, might have a primordial origin
\cite{dynamo}. If this is true, then the plasma of the early
universe sustained enormous magnetic fields which may have
affected also particle interactions. It is natural to assume that
the primordial field is random. This is because any magnetic
field is imprinted on the almost chaotically comoving plasma,
which entangles and mixes the field lines.
 Recently it was pointed out\cite{SemikozValle}  that in
presence of such random  magnetic fields  the BBN constraints on
sterile neutrino become more stringent than in isotropic plasma.

In the present paper we investigate the kinetics of   the
active-sterile neutrino conversions in  hot plasma of the early
universe using a set of quantum kinetic equations.
(For earlier applications, see\cite{early,EKT,Rudzsky}.)
 Our formalism is general and takes also into account
 quantum damping as well as the effects of randomly magnetized  plasma.
 Apart from reproducing the earlier results obtained in
particle approach, the kinetic approach reveals new effects, which give rise
to a novel constraint on the neutrino mixing parameters.
It turns out to be more stringent than the previous ones, which have been
obtain
ed from one-particle Schr{\"o}dinger equation.

The organization of the paper is as follows. To
describe   neutrino propagation in hot plasma we  derive,
in Section 2,   neutrino
dispersion relations in magnetized plasma. We also discuss briefly
the behaviour of random magnetic fields.
 In Section 3, starting from the quantum kinetic equations (QKE) first
 derived in\cite{Thomson1},
we obtain  a general  analytic solution for the  neutrino
conversion probability in equilibrium plasma at   temperatures
$T\gg m_e$. Special care is taken to include correctly the
effects of a possible random backround magnetic field.
In Section 4 we apply our general formalism to the special case of
isotropic plasma. We find a new
shrinking,  aperiodic analytic solution for $\nu_e\to \nu_s$
neutrino conversions,
 which gives rise to a new, stringent constraint on the mixing
parameters.
 In Section 5 we include a primordial magnetic field and find
for large fields
 a  simple analytic expression for
the conversion  probability. The solution
 is a generalization of the one-particle result of\cite{SemikozValle}.

In Section 6 we discuss the results and their applications both
to  neutrino  physics and to the  astrophysics of galactic
magnetic fields.

\section{Neutrino propagation in hot plasma}

\subsection{Random magnetic fields}
In order to describe
active-sterile neutrino oscillations in a general medium, which may
also contain a random
magnetic field, we need neutrino dispersion relations. We also
need a model for the primordial magnetic field. Although
there are a number of suggestions as to how a large magnetic
field could arise in the early universe\cite{btheories,savvidy},
we prefer a more phenomenological approach.

A primordial magnetic field, no matter what  its origin, is
imprinted on the comoving plasma, which in the early universe
has a very large conductivity.
We shall assume that the primordial plasma consist of
elementary magnetic domains of size $L_0$. Within each such domain the
magnetic field is taken to be uniform and constant, and the
field in different domains is randomly aligned.
For the root mean
squared  magnetic  field $B_{rms} =
\sqrt{\VEV{{\bf B}^2}}$,  averaged over a volume $L^3\gg L_0^3$,
we assume the scaling law
\be[rmsfield]
 B_{rms} = B_0\Bigl (\frac{T}{T_{0}}\Bigr )^2
\Bigl (\frac{L_0} {L}\Bigr )^p~.
\ee

The temperature dependence of $B_{rms}$ reflects  simply magnetic
flux conservation. How does $B_{rms}$ scale
with distance is an unsolved
issue and reflects our ignorance e.g. on the question as to how
uncorrelated the fluxes in the neighbouring magnetic domains actually
are. In the case of no correlation statistical averaging gives
for the parameter $p$ the value $p=1/2$, whereas a random-walk
argument yields the scaling $p=3/2$\cite{rez}.

Dissipation of $B$ was investigated in\cite{rez,Schramm2}.
There it was found that
the dissipation length  at
the recombination time is about
$ 10^{10}~{\rm cm}$, which  corresponds to
$10^{10}~{\rm cm}\times (T_{rec}/T_{BBN})
\simeq 2\times 10^4~{\rm cm}$ at nucleosynthesis.
If primordial magnetic fields are to be the seed field for the galactic
dynamo\cite{dynamo}, the primordial field should survive until recombination.
This sets a limit for the size of the random magnetic field
domain $L_0$:
 \be[domainsize]
 L_0\geq L_0^{min} = 10^3~{\rm cm}\times \Bigl(
{\MeV\over T}\Bigr )~.
\ee
Primordial nucleosynthesis may be used to set limits on the size of $B$
\cite{Schramm2,blimits}.

For an uncorrelated random magnetic field  the mean field
$\VEV{B_j} = 0$, whereas

\be[rule]
\VEV{B_i({\bf x})B_j({\bf x}')} =
\frac{1}{2\lambda}\delta_{ij}\delta^{(3)} ({\bf x} - {\bf x}')~.
\ee
Here the correlation length $\lambda$ is determined by the domain size
$L_0$ and the value of root mean squared field
$B_{rms}$ at the horizon scale $L=l_H(T)$ \cite{rez}:

\be[lambda]
\begin{array}{cc}
\frac{1}{\lambda} = \frac{3}{\pi
(3-2p)}B_{rms}^2(l_H)L_0^3~,~~p\neq 3/2~,\\
\frac{1}{\lambda} = \frac{3}{\pi}\Bigl (\ln \frac{l_H}{L_0}\Bigr
)B_{rms}^2(l_H)L_0^3~,~~p = 3/2~.
\end{array}
\ee
\noi
The horizon scale defines a cut-off $k_{max}$ for  the wave number.
Physically it corresponds to the scale of
the inhomogeneity, $k_{max}=2\pi/L_0$\cite{rez}. This is the reason
for the appearance of the horizon size in the above formulas.

\subsection{Neutrino dispersion relations}

The ultrarelativistic dispersion of a standard electroweak neutrino
in hot plasma depends on the neutrino
interaction potential, which consists of two parts:

\be[potential]
 V = V^{(vec)} + V^{(axial)}.
\ee
The potential is determined by the neutrino forward scattering amplitude
off all particle species in the plasma, including magnetized
charged leptons and antileptons.  The vector interaction
potential
$V^{(vec)}$ for a neutrino with a momentum $p=\langle p\rangle =3.15~T$ is
given by\cite{Notzold}

\be
V^{vec}=G_F\sqrt{2}n_{\gamma}[n_{\rm asym} - A\frac{T^2}{M_W^2}],
\ee
where $n_{\rm asym}$ depends on the particle-antiparticle asymmetries
in the plasma, normalized to the photon density
$n_{\gamma}=0.244 T^3$, and
$A\simeq 55$. In the hot primordial plasma with $T\gg
m_e$, where particle asymmetries are small, the vector interaction
potential $V_{\rm vec}$ is dominated by the second, non-local
term, which reads

\be[vector]
|V^{(vec)}|\simeq 3.4\times 10^{-20}\Bigl
(\frac{T}{\rm MeV}\Bigr )^5~{\rm MeV}~.
\ee

\noi
 The axial potential $V_{\rm axial}$ is present only if
the plasma supports a magnetic field.
It is given  by\cite{SemikozValle}

\be[axial]
 V^{(axial)} = \mu_{eff}\frac{{\bf k\cdot B}}{k} +
\frac{\mu_{eff}^2}{2k}\Bigl ( B^2 - \frac{({\bf
k\cdot B})^2}{k^2}\Bigr )~.
\ee
\noi
Here the quantity $\mu_{eff}$ is defined by
\be
\mu_{eff} = \frac{eG_F(-2c_A)T\ln 2}{\sqrt{2}\pi^2}\simeq
-12~c_A
\times 10^{-13}\mu_B\Bigl (\frac{T}{\MeV}\Bigr )~,
\ee

\noi
where $c_A = \mp 0.5$ is the axial coupling in the weak
lepton current  (the upper sign is for $\nu_e$, the lower
one for $\nu_{\mu,\tau}$), and  $\mu_B = e/2m_e$ is the  Bohr
magneton. In our case it is sufficient to consider only the
first term of $V^{(axial)}$ since according to eq.
\eq{rmsfield}
$\mu_{eff}B\ll  k\sim 3T$.

\section{Neutrino conversions in hot plasma}

\subsection{Averaging over random magnetic fields}

Quantum kinetic equations  take
into account the inherent quantum nature of neutrino
oscillations (\cite{Thomson1} and references therein, see
also\cite{Peletminsky}).
Adopting this approach we now derive  a general
expression for  the probability of active-sterile neutrino
conversions in the early universe.

The time evolution of the  system of the active neutrino
$\nu_e$ and a sterile neutrino $\nu_s$ can be described in terms
of a  polarization vector $(P_0(t),~{\bf P}(t))$, whose
z-component gives  the the excess of $\nu_s$ over $\nu_e$
in the neutrino
ensemble at a given moment of time.  In Ref.\cite{Thomson1} the
general time evolution equations  of the  polarization vector
were derived (eqs. (28) to (31) of
\cite{Thomson1}), and these are applicable for any neutrino transitions.
In our case,
 where one of the neutrinos is sterile with  no interaction with
the background matter, these  equations  are
considerably simplified.

We shall assume a hierarchy between the  relaxation times  of the
spatial distribution along the neutrino trajectory, and the
momentum  distributions.
 This assumption allows us to factorize out
 the equilibrium momentum distribution
$f_{\nu}(k)= [\exp (k/T) + 1]^{-1}$, so that the  density matrix
can be written as ($\hat{\bf k}$ is the unit vector in the direction of ${\bf
k}$)
\be[separation]
\rho_{\nu}(t,{\bf k}) = \frac{1}{2}[P_0(t) + P_i(t,
\hat{\bf k})\sigma_i] f_{\nu}(k)~.
\ee

\noi
The distributions of the active neutrino
$\nu_e$ and the  sterile neutrino $\nu_s$ can be written as
\be[distribution]
\begin{array}{cc} f_{\nu_e}(t, {\bf k}) = \frac{1}{2}[P_0(t) +
P_z(t,
\hat{\bf k})]f_{\nu}(k)~,\\ f_{\nu_s}(t, {\bf k}) =
\frac{1}{2}[P_0(t) - P_z(t,
\hat{\bf k})]f_{\nu}(k)~.
\end{array}
\ee
Let us note that our definitions \eq{separation} and \eq{distribution} are
more general than those given in\cite{Thomson1}.

There is a   hierarchy of scales in our problem:
$L_0\ll L_W\ll l_H$, where $L_0$ is the previously defined
scale of elementary domains in the randomly distributed magnetic
field,
$L_W$ is the length scale of weak interactions and $l_H$ is the
horizon scale. Therefore we should  average our QKE  over the
random magnetic field distribution  before integration   over
the  momenta.
 Moreover, we are interested in the neutrino
conversion probability at  physical distance much larger than
$ L_0$. It is described by the  mean value
 of the flavour space polarization vector
$\VEV{P_z(t, \hat{\bf k})} = P_z(t)$, which does not depend on the
neutrino momentum.

As the temperature changes slowly as compared with the
 neutrino oscillation rate one can write, using  factorization
\eq{distribution}, the active-sterile  evolution equations
in the form\cite{Thomson1}
\beqa
\frac{dP_0(t)}{dt} &=& R(t,k)/f_{\nu}(k)~,\label{QKE4}\\
\frac{dP_x(t, \hat{\bf k})}{dt} &=& - V_z(t, \hat{\bf k})P_y(t, \hat{\bf k})
-  D(t,k)P_x(t, \hat{\bf k})~,\label{QKE1}\\
\frac{dP_y(t, \hat{\bf k})}{dt} &=& V_z(t, \hat{\bf k})P_x(t,
\hat{\bf k}) - P_z(t, \hat{\bf k}) V_x - D(t,k)P_y(t,
\hat{\bf k})~,\label{QKE2}\\
\frac{dP_z(t, \hat{\bf k})}{dt} &=& V_xP_y(t, \hat{\bf k}) + R(k,
t)/f_{\nu}(k)~.\label{QKE3}
\eeqa
Here $R(k,t)$ is the rate of the annihilation processes
($\nu_e\overline{\nu}_e\leftrightarrow e^+e^-$), the general
form of which is given in\cite{Thomson1}. The
damping parameter $D(k,t)$ is determined by the inelastic- and
elastic  neutrino collisions.

The neutrino interaction potential $V_z$ is given by
\be[nupot]
 V_z(t, \hat{\bf k}) = V^{(vec)} -
\Delta \cos 2\theta_0 + \mu_{eff}B_{\parallel}(t)
\ee
where $V^{vec}$ is given by the eq. \eq{vector}, the magnetic
term is approximated from eq. \eq{axial}, $\Delta = (m^2_1 - m^2_2)/2k$, and
$V_x=\Delta\sin 2{\theta_0}$.

 The system of eqs.
 \eq{QKE4} yields the following quite complicated
integro-differential equation  for $P_z$:
\beqa
 &&[\ddot{P}_z(t) + V_x^2P_z(t) + (\dot{P}_z(t) -
{R(t)\over f_{\nu}})D(t)]F(t)
 \nonumber\\
&&+ (V^{(vec)} - \Delta \cos
2\theta_0)^2\int_0^t dt_1
\Bigl[\frac{dP_z(t_1)}{dt_1} - {R(t_1)\over f_{\nu}}\Bigr]F(t_1)
 \nonumber\\
&&+
\mu_{eff}^2\int_0^t dt_1 B_{\parallel}(t)B_{\parallel}(t_1)\Bigl
[\frac{dP_z(t_1)}{dt_1} - {R(t_1)\over f_{\nu}}\Bigr]F(t_1)\nonumber\\
 &&+
(V^{(vec)} - \Delta\cos
2\theta_0)\mu_{eff}\Bigl\{B_{\parallel}(t)\int_0^t dt_1
\Bigl (\frac{dP_z(t_1)}{dt_1} -
{R(t_1)\over f_{\nu}}
\Bigr)F(t_1)\nonumber\\
&&+ \int_0^t dt_1 B_{\parallel}(t_1)\Bigl (\frac{dP_z (t_1)}{dt_1} -
R(t_1)/f_{\nu}\Bigr)F(t_1)\Bigr\}\nonumber\\ &&= 0~,
\label{intdif}\eeqa
where $F(t)=\exp(\int_0^{t}D(t')dt')$
(here we  omit  in all the
 arguments the momentum variable ${\bf k}$.) In the  case of a
random magnetic field  integration  over the
dimensions transversal to the neutrino pragation direction
allows us to omit the  terms  linear in $B_{\parallel}$.
For the product
$B_{\parallel}(t)B_{\parallel}(t_1)$ we may use the
approximate result\cite{SemikozValle}
$\VEV{B_{\parallel}(t)B_{\parallel}(t_1)}/\VEV{B_{\parallel}^2}
\sim L_0\delta (t - t_1)$, so that this term reduces to
\be
 2\Gamma\Bigl[\frac{dP_z(t)}{dt} - R(t)/f_{\nu}
\Bigr]\exp (\int_0^tD(t_2)dt_2)~.
\ee
 Here $\Gamma$ is the damping induced by  the presence of a
magnetic field. Using  eq. \eq{lambda} for $p\neq 3/2$ yields
 the following expression:
\be[damping]
\Gamma= \mu_{eff}^2
\VEV{B_{\parallel}^2}L_0=
\frac{3}{4\pi(3 - 2p)}\mu_{eff}^2B_{rms}^2(L=l_H)L_0~.
\ee

Differentiating eq. (\ref{intdif}) with respect to time, and omitting the
common exponential factor, we finally obtain the following third order
differential equation:

\beqa
 &&f(k)\Bigl [\frac{d^3P_z}{dt^3} + 2\bigl(\Gamma +
D(t,k)\bigr)\frac{d^2P_z}{dt^2}\nonumber\\
&&+ \left(\Delta_m^2 + 2\Gamma D(t, k) +
D^2(t,k) +
\frac{dD(t, k)}{dt}\right)\frac{dP_z}{dt} + D(t,k)V_x^2P_z\Bigr ]
\nonumber\\
&&= R(t,k)\Bigl [2\Gamma D(t, k) + D^2(t,k) + \frac{dD(t,
k)}{dt} + (V^{(vec)} -
\Delta \cos 2\theta_0)^2\Bigr ]
\nonumber\\
&&
+ 2\Gamma \frac{dR(t,k)}{dt}~,
\label{intermediate}\eeqa
 where $\Delta_m$ is the usual neutrino oscillation frequency
in isotropic  medium:
\be[MSW]
\Delta_m = [(V^{(vec)} - \Delta \cos 2\theta_0)^2 + V_x^2]^{1/2}~.
\ee

\subsection{Master equation for active-sterile conversion}

 Before  neutrino decoupling, pair  annihilation  does not
contribute to the factor $R$,  but it contributes, like
elastic neutrino  collisions, to the collision damping
coefficient $D(t,k)$. At fixed temperature $D$ does not depend on
time. This allows us  to rewrite the evolution equation of the
active-sterile  neutrino conversion probability

\be[probability]
P  = P_{\nu_e\rightarrow \nu_s}=
 \frac{1}{2}\left[1 - \frac{P_z(t)}{P_0(0)}\right]
\ee

\noi
in the
following  simplified form:
\be[differ]
\frac{d^3P}{dt^3} + 2(\Gamma + \Gamma_W)\frac{d^2P}{dt^2} +
(\Delta_m^2 + 2\Gamma\Gamma_W + \Gamma_{W1}^2)\frac{dP}{dt} +
\Gamma_WV_x^2P = \frac{
\Gamma_WV_x^2}{2}~.
\ee
  Here we have  averaged  eq. \eq{intermediate} over
the normalized momentum distribution,  and  used the
notations
\beqa
\Gamma_W &=& \int dkf(k)D(k)=\langle D\rangle,\nonumber\\
\Gamma_{W1}^2 &=& \int dkf(k)D^2(k)= \langle D^2\rangle~,
\eeqa
where $ dk = \Omega d^3k/(2\pi)^3$.

Eq. \eq{differ} is our master equation for the active-sterile
neutrino conversion. The initial conditions are easily determined
from the equations
\eq{QKE4}--\eq{QKE1} and are given by
\be[initial] P(0) = 0~,~~
\dot{P}(0) = 0~,~~
\ddot{P}(0) = \frac{V_x^2}{2}~.
\ee
The solution of eq. \eq{differ} is of the
general form
\be[solution] P(t) = \frac{1}{2} + C_1e^{k_1t} + C_2e^{k_2t} +
C_3e^{k_3t}~.
\ee
 Here $k_{1,2,3}$ are the roots of the Cartan equation
\be k^3 + a_2k^2 + a_1k + a_0 = 0~,
\ee
where the constant coefficients read
\be[aconstant]
\begin{array}{ll} a_2 = 2(\Gamma + \Gamma_W)~,\\
 a_1 = \Delta_m^2
+ 2\Gamma\Gamma_W + \Gamma^2_{W1}~,\\
 a_0 = \Gamma_WV_x^2~.
\end{array}
\ee
The roots
\be[k-roots]
\begin{array}{ll} k_1 = - \frac{2(\Gamma + \Gamma_W)}{3} + (s_1 +
s_2)~,\\
 k_2 = - \frac{2(\Gamma + \Gamma_W)}{3} - \frac{1}{2}(s_1
+ s_2) + \frac{i
\sqrt{3}}{2}(s_1 - s_2)~,\\
 k_3 = - \frac{2(\Gamma +
\Gamma_W)}{3} -
\frac{1}{2}(s_1 + s_2) - \frac{i
\sqrt{3}}{2}(s_1 - s_2)~,\\
\end{array}
\ee
can be found by solving the Cartan equation in the standard way so
that
$s_1=(r+\sqrt{r^2+q^3})^{1/3},~s_2=(r-\sqrt{r^2+q^3})^{1/3}$ with
$r=(a_1a_2-3a_0)/6-a^3_2/27,~q=a_1/3-a_2^2/9$.
 The constants   $C_i = C_i(k_1, k_2, k_3)$ in eq. \eq{solution} are
determined from the initial conditions
\eq{initial}.

In the general case the roots $k_i$ are quite
complicated, but in some special  cases they reduce to a rather simple
expression. For instance, in the absence of collisions one has
$\Gamma_W = \Gamma^2_{W1} = 0$, so that eq. \eq{differ} can easily be
integrated. The probability eq. \eq{probability} takes in this case
the form:
\be[old] P(t) = \frac{V_x^2}{2\Delta_m^2}\Bigl \{1 - \exp
(-\Gamma t)\Bigl [\cosh(
\sqrt{\Gamma^2 - \Delta_m^2}t) + \frac{\Gamma}{\sqrt{\Gamma^2 -
\Delta_m^2}}
\sinh(\sqrt{\Gamma^2 - \Delta_m^2})\Bigr ]\Bigr \}~.
\ee
This result, which here followed from the general kinetic approach,
was previously derived in\cite{SemikozValle} by using a different method.
 One can also readily see from eq. \eq{old} that in the absence of a
 magnetic field our approach gives rise to the standard MSW
result for the neutrino conversion  probability\cite{MSW}:
\be[standard] P(t) =
\frac{V_x^2}{\Delta_m^2}\sin^2\frac{\Delta_mt}{2}~.
\ee
\section{Shrinkage of the flavour spin}

\subsection{Solution for $B=0$}

Let us first
apply our general formalism to the simple case of vanishing magnetic
field  by
setting $\Gamma=0$.  This has been previously studied in the literature
by using the one-particle approach. The new effect revealed by the quantum
kinetic approach is the shrinkage of the flavour polarization
vector, which results in a more stringent bound on the vacuum oscillation
parameters than the previous ones. The effect was first
discussed  qualitatively by Stodolsky in\cite{Stodolsky}.

We take
$\Delta_m \simeq V^{(vec)}\gg \Gamma_W~$, i.e. we assume that the system is far
from the MSW resonance. We then find
\be s_1 = \frac{\Delta_m}{\sqrt{3}}\Bigl (1 +
\frac{\Gamma_W\sqrt{3}}{3\Delta_m} + \frac{\Gamma_{W1}^2 -
\Gamma_W^2}{2\Delta_m^2} -
\frac{\Gamma_WV_x^2\sqrt{27}}{ 6\Delta_m^3} + O\Bigl
((\Gamma_W/\Delta_m)^4\Bigr )~,
\ee
\be s_2 = - \frac{\Delta_m}{\sqrt{3}}\Bigl (1 -
\frac{\Gamma_W\sqrt{3}}{3\Delta_m} + \frac{\Gamma_{W1}^2 -
\Gamma_W^2}{2\Delta_m^2} + \frac{\Gamma_WV_x^2\sqrt{27}} {
6\Delta_m^3} + O\Bigl ((\Gamma_W/\Delta_m)^4\Bigr )\Bigr )~.
\ee
We thus have $s_1 + s_2 = 2\Gamma_W/3 -
\Gamma_W(V_x^2/\Delta_m^2)$ and
$s_1 - s_2 = (2/\sqrt{3})\Delta_m(1 + (\Gamma^2_{W1} -\Gamma_W^2)/
2\Delta_m^2)$, where in the difference we can neglect the
dispersion  term
$\VEV{D^2} - \VEV{D}^2 = \Gamma_{W1}^2 - \Gamma_W^2$.
 Substituting $s_1$ and $s_2$ into the roots $k_i$ given by eq.
\eq{k-roots}, we obtain $k_1 = - \Gamma_W(V_x/\Delta_m)^2$,~~
$k_2 = -
\Gamma_W  + i\Delta_m$ and $k_3 = - \Gamma_W - i\Delta_m$. The
transition  probability \eq{solution} is then given by
\be[new1] P_{\nu_e\rightarrow \nu_s} = \frac{1}{2}\Bigl \{1 -
\exp \Bigl (-(V_x/\Delta_m)^2\Gamma_Wt\Bigr )  +
\frac{V_x^2}{\Delta_m^2}\Bigl [\exp \Bigl (-(V_x/
\Delta_m)^2\Gamma_Wt\Bigr ) - e^{-\Gamma_Wt}\cos \Delta_mt\Bigr
]\Bigr \}~.
\ee
For times much larger than the collision time
$\Gamma_W^{-1}$,
 but  much less than the  depolarization time
\be[depolarization] t_d = \frac{\Delta_m^2}{V_x^2}\Gamma_W^{-1}~,
\ee
 neutrino harmonic oscillations die away and the probability for
$\nu_e
\to \nu_s$ conversion is aperiodic:
\be[new2]
P = \frac{1}{2}\Bigl [1 - \exp \Bigl (-
\frac{t}{t_d}\Bigr )\Bigr ]\simeq
\frac{t}{2t_d}~.
\ee
This probability is small,
\be[new3] P \simeq \frac{V_x^2}{2\Delta_m^2}\Gamma_Wt \ll 1~,
\ee
but nevertheless the aperiodic term dominates over the contributions
from the last two terms
in the brackets in eq. \eq{new1}.

The earlier result for the transition probability\cite{EKT} reads,
\be[stan]
  P = V_x^2/2\Delta_m^2 = \frac{\sin^22\theta_0}{2(1
- 2x\cos 2\theta_0 + x^2)}~,
\ee
where  $x$ is defined through
\be[temperature]
\frac{T}{\MeV} = \Bigl (\frac{10^7\times x\Delta m^2}{2~\eV^2}\Bigr
)^{1/6}~.
\ee
As can be seen,
our result eq.  \eq{new2}
differs from  eq. \eq{stan}
 by an additional large factor $\Gamma_Wt\sim \Gamma_W/H\sim
(T/{\rm MeV})^3$. This is due to spin shrinkage which is manifest in the
kinetic approach.
\subsection{Heuristic derivation of spin shrinkage}

 The result eq. \eq{new2} has been discussed earlier by
Stodolsky\cite{Stodolsky}, who  also pointed out that a  large conversion
 probability $P=1/2$ may never be reached because
depolarization time $t_d$ could be very large, in our case larger
than the age of universe. Depolarization time
can be estimated in a simple manner by noting that
  flavour spin rotation turns the longitudinal part of the spin
into transversal, and at each collision the transversal part is,
in effect, wiped out. This results in a shrinkage of the spin
vector, and as
$t\to\infty$, $P_{\nu_e\to\nu_s}(t)$ approaches  the value 1/2.
The length of the spin vector ${\bf P}$ gives the
degree of coherence, and as $t\to\infty$, ${\bf P}$ vanishes, which
corresponds to a completely mixed incoherent state.

Qualitatively one can see this by writing
 the QKE \eq{QKE4}--\eq{QKE3} in the familiar form\cite{Stodolsky}
\be
\frac{d{\bf P}}{dt}={\bf V}\times{\bf P} - D{\bf P}_{\perp}~,
\ee where now ${\bf V}=V^{(vec)}\hat{n}_z + V_x\hat{n}_x$, $D
\sim \Gamma_W$ and the transversal spin is given by $P_\perp\sim
P(V_x/V_z)$ with $V_x =
\Delta \sin 2\theta_0$. It then follows that
\be {dP^2\over dt}=-2\Gamma_W P^2_{\perp}~,
\ee yielding a shrinkage rate $P^{-2}dP^2/dt= - \Gamma_W\times
V_x^2/(V^{(vec)})^2$,  in agreement with eq. \eq{new2}. Thus, in
the presence of collisions the one-particle matrix density
\be
\rho_{\nu}(t, k) = \frac{1}{2}[P_0(t) +
P_i(t)\sigma_i]f_{\nu}(k)
\ee tends in the formal limit $t\to \infty$ to the unpolarized
one because of the slow spiralling of the spin vector into the origin:
\be
\rho_{\nu}(t,k)\to \frac{1}{2}P_0(\infty)f_{\nu}(k)~.
\ee
 The outcome is a unique equilibrium point for the system, given
by $P_x = P_y = P_z = 0$,  where the number of active neutrinos
equals the number of sterile neutrinos.

\subsection{Big bang nucleosynthesis constraint}
The aperiodic term  modifies considerably the constraint on the
neutrino mixing
parameters. Let us look at this matter more closely by first recapitulating the
derivation of the earlier result\cite{EKT}, and then presenting a similar
deriva
tion within our present
approach.

The condition for non-equilibration of sterile neutrinos is
 $\Gamma_s=P\Gamma_W\lsim H$, which implies the constraint
\be[BBN3]
\Bigl (\frac{\Gamma_s}{H}\Bigr )_{max} = 0.6\sin^22\theta_0\times
\sqrt{ 10^7\mid \Delta m^2\mid/(2~\eV^2)}f_1^{max}(y)\lsim1~,
\ee
where $y = \pm x$ depending on the sign of the  mass difference
$\Delta m^2 = m_1^2 - m_2^2$. The function $f_1(y)$, defined as
\be[function1]
f_1(y) = \frac{\sqrt{y}}{1 + 2y\cos 2\theta_0 +
y^2}~,
\ee
is  the product of the probability \eq{stan} and the ratio
$\Gamma_W/H$. For a small mixing angle it reaches the
maximum value
$f_1^{max}\simeq 3\sqrt{3}/16$ at $y= (\sqrt{3 +
\cos^22\theta_0} - \cos 2\theta_0)/3$.
 Substituting these values into eq. \eq{BBN3} one obtains the
following allowed region for the mixing parameters\cite{EKT}:
\be[oldlimit]
\mid \Delta m^2\mid \sin^42\theta_0 \lsim 5\times 10^{-6}~\eV^2~.
\ee
For a typical mixing angle $\sin 2\theta\simeq 0.1$ this yields
$
\mid \Delta m^2\mid \lsim 5\times 10^{-2}~\eV^2~.
$

In our case, where the conversion probability is given by eq.
\eq{new3}, one obtains the
constraint
\be[BBN2]
\Bigl (\frac{\Gamma_s}{H}\Bigr )_{max} = 2.5\times
10^6\sin^22\theta_0\times
\frac{\mid \Delta m^2\mid}{eV^2}f_2^{max}(y)\lsim 1~,
\ee where the function
\be[function2] f_2(y) = \frac{y}{1 + 2y\cos 2\theta_0 + y^2}
\ee
has the maximum value $f_2^{max}=1/ 4\cos^2\theta_0$ at $y=
1$.
 Thus, in our case
 the allowed region is
\be[newlimit]
\mid \Delta m^2\mid\sin^22\theta_0\lsim 1.6\times 10^{-6}~{\rm eV}^2~.
\ee

One should notice the  different $\sin 2\theta_0$ dependence compared with
the earlier result \eq{oldlimit}. This difference gives rise to a
more stringent constraint on  the neutrino mass in our case. For example, for
$\sin 2\theta_0\sim 0.1$ we find
$
\mid \Delta m^2\mid \lsim 1.6\times 10^{-4}~{\rm eV}^2~.
$
 The old and the new constraints in the ($\Delta
m^2~,~\sin 2\theta_0$)-plane are depicted in the Figure.

\section{Constraints in the presence of a magnetic field}

\subsection{Conditions for large magnetic effects}

Let us now consider neutrino propagation in hot
magnetized plasma as described by the
master  equation \eq{differ}. In this section we  will assume a
non-vanishing  magnetic
damping
$\Gamma$ in eq. \eq{differ}, together with a non-vanishing quantum damping
$\Gamma_W$.
We assume that the neutrino system has not yet reached the MSW resonance
region, so that the oscillation frequency in matter $\Delta_m$ can be
approximated by $\Delta_m\simeq  V^{(\rm vec)}$. Furthermore, we take
  $\Delta_m^2\gg
\Gamma_W^2~,~{\Gamma_{W1}^2}$. We also assume that
the dominant plasma effect  on neutrino
prapagation  is  due to random magnetic fields as given by eq. \eq{rmsfield}.
This is the case where the magnetic damping rate is much larger than the
oscillation rate of neutrinos, $\Gamma\gg \Delta_m$, and simultaneously
$\Delta_m^2/\Gamma\Gamma_W\ll 1$. With these approximations, we are able
to solve the QKE \eq{differ} analytically.

Let us study more closely these approximations.
We may note that the magnetic damping rate
\be
\Gamma = 3\mu_{eff}^2B_{rms}(L = l_H)L_0/4\pi(3 - 2p)~
\ee  has a minimum value
\be[gammamin]
\Gamma\geq \Gamma_{min} = \frac{3.5(2.3)^{2p}}{\pi(3 - 2p)}
\times 10^{-12 - 16p}{B^2_{rms}(L = l_H)\over 10^{24} G}
\Bigl (\frac{T}{\rm MeV}\Bigr )^{5 + 2p}~\left({B_0\over T_0^2}\right)^2{\rm
MeV}
\ee
which corresponds to the minimum value of
the magnetic domain size $L_0^{min}$, given in \eq{domainsize}.

Let us now make a further assumption. To illustrate the potential importance
of the magnetic effects, we take $B_0\simeq T_0^2$, which we believe is
the largest possible random field that can be supported by the plasma.
An example of this kind of a situation is possibly provided by the
electroweak phase transition, where it has been argued\cite{btheories}
that  fluctuating Higgs field
gradients could generate magnetic fields of the order of $T_{EW}^2$.

To see when magnetic effects are important,
we  compare magnetic damping with $\Gamma_W\simeq  4.0G_F^2T^5$ and
the oscillation frequency $\Delta_m\simeq  V^{(\rm vec)}$. Since
$V^{(\rm vec)}\gg \Gamma_W$, a sufficient condition for the domination
of magnetic damping is $\Gamma_{min}>V^{(\rm vec)}$. It turns out that
even more restrictive condition is  the requirement
$\Delta_m^2/2\Gamma_{min}\Gamma_W\ll 1~.$
This condition implies the following range of  validity of the approximations
discussed above:
\be[temperature1]
\frac{T}{\rm MeV}> \frac {(3\pi(3 - 2p))^{1\over 2p}}{2.3}\times 10^{8 -
{7\over 2p}}~.
\ee

We note in passing that if we discard the  hypothesis of the
relic origin of the galactic seed field,  thereby allowing considerably
smaller domain sizes than given in  eq.
\eq{domainsize}, the conditions $\Gamma\gg \Delta_m\gg \Gamma_W$,
$\Delta_m^2/2\Gamma\Gamma_W\ll 1$ can be fulfilled only for very small values
of the  index
$p$, i.e. when the magnetic field is very close to a constant. Our analysis
does not cover this case.

\subsection{Application to primordial nucleosynthesis}

Assuming  the validity of the  approximations discussed in the previous
subsection, we obtain for the roots
\eq{k-roots} the following simple expressions:
\be[simpleroots]
\begin{array}{ll}
 k_1 \simeq - 2\Gamma - \Gamma_W +
\Delta_m^2/2\Gamma\simeq -2\Gamma~,\\
k_2 \simeq - \Gamma_W -
\frac{\Delta_m^2}{2\Gamma}\simeq - \Gamma_W~,\\
k_3 = -
\frac{V_x^2}{2\Gamma} + \frac{\Delta_m^4}{16\Gamma^2\Gamma_W}
\simeq -\frac{V_x^2}{2\Gamma}~.
\end{array}
\ee
In  $k_3$  we have neglected the second term
$\Delta_m^4/16\Gamma^2
\Gamma_W$ as it is resonable to assume that  $V_x\leq \Delta_m$.
 From eq.
\eq{simpleroots} one can then easily see that for  $t\gg
\Gamma_W^{-1}$ only the last term in eq.
\eq{probability}, the one proportional to  $ C_3e^{k_3t}$, survives, so that
the active-to-sterile neutrino conversion probability now takes  the form
\be[simple]
P \simeq \frac{1}{2}[ 1 - e^{-V_x^2t/2\Gamma}]~.
\ee

In order to avoid a conflict with the primordial nucleosynthesis
constraint, the production rate $\Gamma_s=P\Gamma_W$ of the sterile neutrinos
should be smaller than the expansion rate $H\simeq t^{-1}$
of the universe in the temperature range $1~\MeV\lsim T\lsim T_{\rm QCD}$.
When $T\lsim 1$ MeV, left chiral neutrinos are decoupled, and collisions
can no longer excite sterile neutrinos. Above the QCD phase transition
temperature $T_{\rm QCD}$ sterile neutrino production  is harmless,
because their number densities are subsequently suppressed by the
relative heating of
the interacting species. From eq. \eq{temperature1} we find that,
within our approximative solution, nucleosynthesis constraint is relevant only
if $p\lsim 0.6$. If this is so, we obtain from eq. \eq{simple} the
condition
 $\mid V_x\mid \lsim 2H \sqrt{\Gamma_{min}/\Gamma_W}$.
This can be translated into
a constraint on the neutrino  mixing parameters:
\be[constraint]
\frac{\mid \Delta m^2\mid}{\eV^2}\sin 2\theta_0\lsim 8\times
10^{-2 - 8p}\times
\frac{(2.3)^p}{\sqrt{(3 - 2p)}}\Bigl (\frac{T}{\rm MeV}\Bigr
)^{3 + p}~~~(p\lsim 0.6)~.
\ee
Since this limit  is  more restrictive the lower
the temperature is, we substitute into eq. \eq{constraint} the lower bound
of the validity range defined in eq. \eq{temperature1}. In this way we
obtain our final result
\be[final]
\frac{\mid \Delta m^2\mid}{\eV^2}\sin 2\theta_0\lsim 2(3\pi(3-2p))^{{1\over
2p}}\times 10^{{33\over 2}-{21\over 2p}}.
\ee
This result is displayed for $p=0.5, 0.4$ and 0.3, together with
the $B=0$ case, in the Figure.

We may note that the presence of
a magnetic field implies a limit which is more stringent than if
$B=0$. Naturally, the limit is also more model dependent. In particular,
it very much depends on our assumption that the primordial field
is the seed field for the galactic dynamo, which reflects itself
in $\Gamma_{min}$. Moreover, our results are valid only for rather
weak scaling behaviour. For larger $p$, it is likely that the bound
\eq{final} would be much weaker. On the other hand, weak scaling is
needed if primordial fields are to provide the seed field.
%

\section{Discussion and conclusions}

We have considered active-sterile neutrino
conversions in  hot plasma of the early universe. Our starting point is
a general quantum kinetic equation, and
 in the collisionless limit we reproduce the  results of\cite{SemikozValle}.
The great advantage of the kinetic approach
is that one is able to
follow the evolution of the transition probability over any
number of collisions. In particular, it allows us follow the
chain of collisions over time scales of the order of the Hubble
scale. This fact
provided us with a new analytic solution which actually dominates
the active-sterile conversion
probability in isotropic plasma. It is related  to the shrinkage
of the flavour spin.  Such behaviour is the  combined effect of neutrino
oscillations, which  convert longitudinal spin components to
transversal ones, and  simultaneous damping of the transversal
component by collisions. Qualitatively this effect, progressive decoherence,
 was discussed
earlier in\cite{Stodolsky}. Here it was realized explicitly in the
analytic solution \eq{new1}.

An analogous shrinkage regime exists for Dirac neutrino spin-flip in
plasma with a large scale random
magnetic field \cite{Semikoz2}.  The physical reason of the
shrinkage of the total { real} spin there
is the same as in our case.

It is perhaps surprising that the inclusion of a  magnetic  field,
which leads to an  increase of the energy gap between  active and
sterile  neutrino spectra (see eq. \eq{potential}), tightens the
constraint on the neutrino mixing parameters. In fact, the
transition probability does decrease in the presence of a
magnetic field exactly because of the energy gap increase.
 However, in the random field case the averaged conversion rates
depend also on the neutrino squared mass difference $\Delta m^2$ in a subtle
way. While in the isotropic  case there is a saturation of the
conversion probability as a function of $\Delta m^2$
in the presence of a magnetic field there is  a
linear dependence upon $\Delta m^2$.
This behaviour is reflected in the stronger mixing
angle dependence.

 \vskip 1truecm

{\bf Acknowledgements} One of us (V.S.) acknowledges the high
energy physics group of Department of Theoretical Physics and
Research Institute for Theoretical Physics at Helsinki University for
hospitality. We thank Alexander Rez for
fruitful discussions. The work has been supported by the Academy of Finland.

\newpage
\noindent{\Large\bf Figure Caption}\\
\noindent
The nucleosynthesis constraint on the $\nu_e-\nu_s$ mixing parameters.
(a) old result; (b) new result, which includes the shrinkage
 of the spin vector. Shown are also the limits obtained in the presence of
a primordial random magnetic field  with the scaling
parameter $p=0.3, 0.4$ and 0.5 (solid lines). In all cases the
allowed region is below the contour.

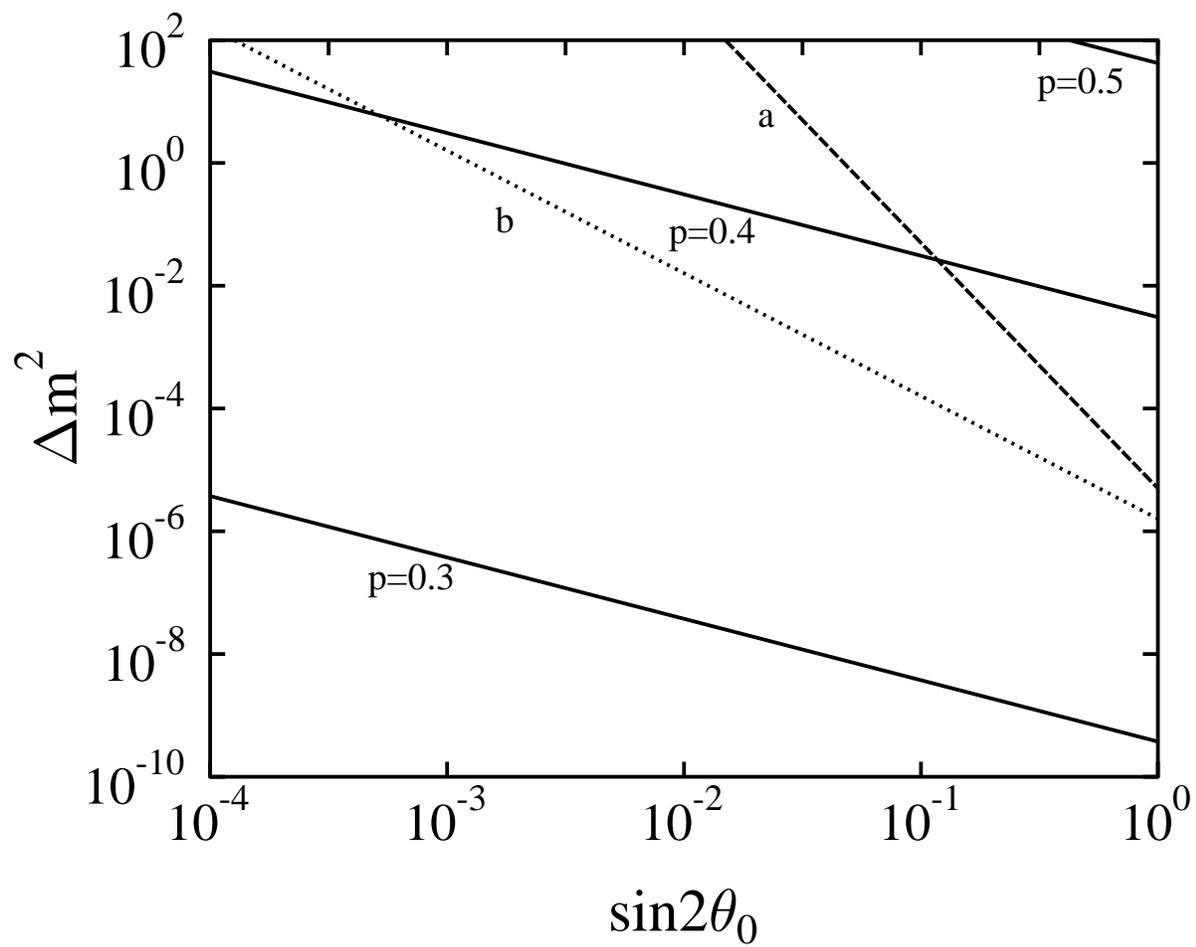
\end{document}